\documentclass[12pt,emtex]{article}

\usepackage{amsfonts}
\usepackage{amssymb}
\usepackage{amscd}
\usepackage{amsmath}

\newcommand{\BEQ}{\begin{equation}}
\newcommand{\EEQ}{\end{equation}}
\def\bea{\begin{eqnarray}}
\def\eea{\end{eqnarray}}
\def\nn{\nonumber}

\newtheorem{Th}{Theorem}
\newtheorem{Lem}{Lemma}

\newtheorem{Rem}{Remark}

\def\bea{\begin{eqnarray}}
\def\eea{\end{eqnarray}}

\def\bes{\begin{equation*} \begin{split}}
\def\ees{\end{split} \end{equation*}}

\def\C{{\mathbb{ C}}}

\def\l{\lambda}

\def\p{\partial_z}

\def\ep{\varepsilon}
\def\xe{\Omega_\ep}

\begin{document}
\begin{titlepage}
\hfill ITEP-TH-54/10

\vskip 3.0cm
\centerline{\LARGE \bf Quantum generic Toda system}

\vskip 1.0cm \centerline{D. Talalaev \footnote{E-mail:
dtalalaev@yandex.ru} }
\vskip 1.0cm \centerline{ MSU \footnote{Lomonosov Moscow State University, GSP-1, Leninskie Gory, Moscow, 119991, Russia.
}, ITEP \footnote{Institute for Theoretical and Experimental Physics, 25 B. Cheremushkinskaya, Moscow, 117259, Russia.}}
\vskip 1.0cm 

\begin{abstract}
The Toda chains take a particular place in the theory of integrable systems, in contrast with the linear group structure for the Gaudin model this system is related to the corresponding Borel group and mediately to the geometry of flag varieties.
The main goal of this paper is to reconstruct a "spectral curve" in a wider context of the generic Toda system \cite{D}. This appears to be an efficient way to find its quantization which is obtained here by the technique of quantum characteristic polynomial for the Gaudin model \cite{T04} and an appropriate AKS reduction.
We discuss also some relations of this result with the recent consideration of the Drinfeld Zastava space \cite{R}, the monopole space and corresponding Borel Yangian symmetries \cite{GKLO}. 
\end{abstract}


\end{titlepage}

\section{Introduction}
The subject of this work is a very particular example of an integrable system - the generic Toda system related to the $A_n$ root system. The method used here is based on the concept of the spectral curve on both classical and quantum levels. The method of the spectral curve and more generally the algebraic-geometric methods in integrable systems provide an intriguingly effective and universal way in describing, solving and quantizing dynamical systems. This work was challenged by the initial construction of the commutative family \cite{D} which is far from the space of spectral invariants for some evolving linear operator.  

Let us remind the spectral curve construction in open and periodic Toda chains due to \cite{Kr}.  The open Toda chain is defined by the Hamiltonian function
\bea
H=\sum_{k=1}^n \frac {p_k^2} 2 +\sum_{k=1}^{N-1}e^{q_k-q_{k+1}}\nn
\eea
and canonical Poisson brackets on variables $p_k, q_l.$
It has the Lax representation with the Lax operator:
\bea
L(w)=\left(
\begin{array}{ccccc}
v_0 & c_0 & 0 & \ldots & 0\\
c_0 & v_1 & c_1 & \ldots & 0\\
\cdot & \cdot & \cdot & \cdot & \cdot \\
0 & \ldots & c_{n-3} & v_{n-2} & c_{n-2} \\
w c_{n-1} & \ldots & 0 & c_{n-2} & v_{n-1}
\end{array}\right).
\eea
where
\bea
c_k=e^{(q_k-q_{k+1})/2},\qquad v_k=-p_k.\nn
\eea
Let us remark that this Lax representation is not unique for the open chain but unifies the spectral curve technique in open and periodic cases. The commutative family is defined by the coefficients of the characteristic polynomial 
\bea 
det(L(w)-\l)=0\nn
\eea
which in turn defines a rational curve. This curve can be interpreted as a limit of a hyperelliptic curve in the periodic case. In fact the open chain is a limit of the system in a quite wider setup - the generic Toda system.

We just outline here the main strategy. We start by introducing a generating function for the classical integrals of the generic Toda system. This function appears a limit of the classical characteristic polynomial for the Gaudin model with a particular choice of magnetic term. Then remarking that this family is invariant with respect to the Borel group action we realize the AKS reduction with respect to the decomposition $\mathfrak{gl}_n=\mathfrak{b}\oplus \mathfrak{so}_n.$ This idea was generalizable to the quantum level. We used the same elements: we considered the quantum Gaudin model with a particular magnetic term, considered its certain limit, demonstrated the invariance of the resulted commutative family with respect to the Borel group action and realized the quantum AKS reduction.

~\\
{\bf Acknowledgments.}
I'm grateful to my colleagues Yu. Chernyakov and G. Sharygin for stimulating discussions on this subject.
This work was partially supported by the fund "`Dynasty"',
RFBR grant 09-01-00239, the grant for the Scientific school support 5413.2010.1., RFBR-Consortium E.I.N.S.T.E.IN 09-01-92437, the work is also    
supported by Ministry of Education and Science of the Russian Federation under contract 14.740.11.0081. 

\section{Spectral curve for the classical system}
\subsection{Definition}
The generic Toda system for the Lie algebra $\mathfrak{gl}_n$ is obtained in terms of the so-called chopping procedure. Let us consider a symmetric matrix $A$ those elements are generators of the Borel subalgebra $\mathfrak{b}=\mathfrak{b}_-$
\bea
A=\sum_{i\le j}(E_{ij}+E_{ji})\otimes e_{ij}\nn
\eea
where $E_{ij}$ are generators of $End(\mathbb{C}^n),$  $e_{ij}$ for $i\ge j$ are generators of the Lie algebra $\mathfrak{b}.$ The matrix coefficients are interpreted as functions on the dual space to the Lie algebra ${\mathfrak{b}}^*$ which is a Poisson space with the Kirillov-Kostant Poisson bracket.
Let us define also the partial matrices $A_k(\l)$ obtained by deleting $k$ right columns and $k$ upper rows of the matrix $A-\l Id.$ By the result of \cite{D} the complete set of roots of all polynomials
\bea
\label{minors}
\Delta_k(\l)=det A_k(\l)=\sum_i I_{k,i}\lambda^i
\eea
 constitutes a commutative family. The alternative way to define this family is by the help of ratios of coefficients of $\Delta_k(\l).$ One can use the fact that the leading term on $\l$ of $\Delta_k(\l)$ is $\Delta_{n-k}(\l)=\Delta_{n-k}.$ Hence one can introduce a family of characteristic polynomials 
\bea
P_k(\l)=\Delta_k(\l)/\Delta_{n-k}(\l), \qquad k=0,\ldots,n/2.\nn
\eea

\subsection{Generating function}
Let us consider the following matrix $A$ corresponding to the complete Lie algebra $\mathfrak{gl}_n$
\bea
A=\sum_{ij}E_{ij}\otimes e_{ij}.\nn
\eea
We need the notation
\bea
\label{xi}
\xe =\left(\begin{array}{cccc}
0&\cdots&0&1\\
\vdots&\ddots&\ep&0\\
0&\ep^{n-2}&\ddots&\vdots\\
\ep^{n-1}&0&\cdots&0
\end{array}\right).
\eea
One can arrange the coefficients of minors \ref{minors} into the generating series. 
\bea
\label{gen_fun_sln}
P(z,\lambda,\ep)&=&det\left(A z^{-1}+\xe
-\lambda Id\right)\nn\\
&=&\sum_k I_k(z,\l,\ep)=\sum_k \ep^{k(k-1)/2}(I_k^0(z,\lambda)+O(\ep))
\eea
where $I_k(z,\l,\ep)$ are homogeneous in $(z^{-1},\l)$ of degree $n-k.$
Then in particular 
\bea
\Delta_k(\l)=I_k^0(1,\l).\nn
\eea
{\Rem \label{comm} This construction demonstrates for example that the minors commute with respect to the Kirillov-Kostant bracket on $S(\mathfrak{gl}_n).$ Indeed, this algebra is a limit case of the Poisson commutative algebra obtained by the argument-shift method or equivalently by considering the corresponding Gaudin model.}

\subsection{AKS scheme}
Let us remind one of the central concept of the integrable systems theory - the Adler-Kostant-Symes scheme. We use here the following variant: let $\mathfrak{g}$ be a Lie algebra  represented as the direct sum of two subalgebras $\mathfrak{g}=\mathfrak{g}_+\oplus\mathfrak{g}_-.$ The symmetric algebra $S(\mathfrak{g})$ is always considered as an algebra of functions on $\mathfrak{g}^*.$ There is a natural projection map
\bea
i:S(\mathfrak{g})\rightarrow S(\mathfrak{g}_+)\nn
\eea
related to the decomposition of the symmetric algebra
\bea
\label{deco}
S(\mathfrak{g})=S(\mathfrak{g}_+)\oplus \mathfrak{g}_- S(\mathfrak{g}).
\eea
Let us remark that $\mathfrak{g}_- S(\mathfrak{g})$ is a Lie subalgebra.
The map $i$ can be interpreted as the restriction to the space $Ann(\mathfrak{g}_-)\in \mathfrak{g}^*.$ Despite this map is not in general Poisson it preserves in a sense an integrability property.
{\Lem
Let $f,h\in S(\mathfrak{g})$ be invariant with respect to the $\mathfrak{g}_+$ Lie algebra action and commute with respect to the Kirillov-Kostant bracket. Then their images $i(f),~i(h)$ commute with respect to the bracket in $S(\mathfrak{g}_+).$
}
~\\
{\bf Proof.~~ }
Let us consider $f,h\in S(\mathfrak{g})$ decomposed subject to (\ref{deco}):
\bea
f&=&f_+ +f_-,\nn\\
g&=&g_+ +g_-.\nn
\eea
Then
\bea
\{f_-,g_-\}=\{f-f_+,g-g_+\}=\{f,g\}-\{f_+,g\}-\{f,g_+\}+\{f_+,g_+\}=\{f_+,g_+\}.\nn
\eea
Both sides take values in different direct summands of (\ref{deco}) hence vanish.
$\blacksquare$
{\Rem In this section we need a rational generalization for this statement, i.e. the case where the Poisson algebra is the field of rational function on the dual space to the Lie algebra. It is a straightforward generalization and we omit it here. By the way it follows from the statement on the quantum level.}

\subsection{Invariance}
We will show that the ratios $\Delta_k(\l)/\Delta_{n-k}$ are invariant with respect to the Borel subgroup of lower-triangular matrices $B\subset SL(n).$ Let us firstly show that the action of the group on functions $e_{ij}$ can be expressed in terms of the action on the Lax operator $A$
\bea
Ad_g (A):=\sum_{ij}E_{ij}\otimes Ad_g(e_{ij})=\sum_{ij}Ad_{g^T}(E_{ij})\otimes e_{ij}=g^T A ({g^T})^{-1}.\nn
\eea  
The action of the group on the coefficients of the characteristic polynomial is expressed as follows:
\bea
\label{char_def}
det(Ad_{g^T}(A)+\xe z^{-1}-\l)&=&det(A+Ad_{({g^T})^{-1}}(\xe) z^{-1}-\l)\nn\\
&=&\sum_k z^{-k}\ep^{k(k-1)/2}(Ad_g(\Delta_k(\lambda))+O(\ep)).
\eea 

Let us consider an element of the Borel group $g=exp(t e_{j,i})=1+t e_{j,i},~j>i.$  Its action on the matrix $\xe$ is expressed as follows
\bea
({g^T})^{-1}\xe g^T=\xe-t E_{n-i+1,j}\ep^{n-i}+ t E_{i,n-j+1} \ep^{j-1}-t^2 \delta_{j,n-1+1}E_{i,j} \ep^{n-i}.\nn
\eea
This matrix satisfies the property that the lowest term in $\ep$ in each row and in each line is on the antidiagonal. This argue by the way that the lowest term in $\ep$ in the characteristic polynomial (\ref{char_def}) is the same as in the non-deformed one.

Let us now show that the Cartan subgroup acts by a character.  Let us consider an element of $B$ $g=exp(t e_{i,i})$ and its action 
\bea
({g^T})^{-1}\xe g^T&=&\xe+\ep^{i-1}E_{i,n-i+1}(exp(t)-1)+\ep^{n-i}E_{n+1-i,i}(exp(t)-1)\nn\\
&+&\ep^{n-i}\delta_{i,n+1-i}E_{i,i}(exp(t)-1)(exp(-t)-1).\nn
\eea
This observation allows to conclude that the antidiagonal terms of $\xe$ are multiplied by scalars, and this affects the asymptotics  by the following manner
\bea
Ad_g(\Delta_k(\lambda))=\chi_k(g)(\Delta_k(\lambda))\nn
\eea
where $\chi_k(g)$ is the corresponding character.
Hence we have demonstrated the following
{\Th The ratios $(I_{k,i})_+/I_{k,n-k}$ generate a commutative subalgebra:
\bea
[(I_{k,i})_+/I_{k,n-k},(I_{m,j})_+/I_{m,n-m}]=0\nn
\eea
in the field of fractions $\mathcal{F}(S(\mathfrak{b})).$
}

{\Rem
In fact there is a wider invariance in this context. One can consider a series of parabolic subalgebras
\bea
\mathfrak{b}\subset \mathfrak{p}_1\subset\ldots\subset \mathfrak{p}_n=\mathfrak{sl}_n\nn
\eea
such that $\mathfrak{p}_n$ is generated by $\mathfrak{b}$ and positive generators corresponding to roots $\alpha_k,\ldots,\alpha_{n-k-1}.$ Let also consider the series of parabolic groups
\bea
{B}\subset {P}_1\subset\ldots\subset {P}_n={SL}_n.\nn
\eea
}
{\Lem \label{lem_invar} 
\bea
I_{k,i}/I_{k,j}\in \mathcal{F}S(\mathfrak{sl}_n)^{P_k}.\nn
\eea}
This result can be found in \cite{GS}.
{\Rem This commutative family and some extensions with relation to the flag variety geometry is discussed in \cite{CS}.}

\section{Quantization}
The quantum model is constructed  by considering a special limit of the Gaudin commutative algebra related to $3$-point case also known as the argument-shift construction. We also demonstrate that this subalgebra is invariant with respect to the $B$-action on the universal enveloping algebra $U(\mathfrak{sl}_n)$ and provide a quantum analog of the AKS construction which produces a commutative algebra in $U(\mathfrak{b}).$

\subsection{Noncommutative determinant}
Let us consider a matrix $B=\sum_{ij}E_{ij}\otimes B_{ij}$ those elements are elements of some  associative algebra $B_{ij}\in \mathfrak{R}.$ We will use the following definition for the noncommutative determinant in this case 
\bea
det(B)=\frac 1 {n!}
\sum_{\tau,\sigma\in
\Sigma_n}(-1)^{\tau\sigma}B_{\tau(1),\sigma(1)}\ldots
B_{\tau(n),\sigma(n)}.\nn
\eea
There is an equivalent definition. Let us introduce the  operator $A_n$ of the antisymmetrization in $(\C^n)^{\otimes n}$
\bea
A_n v_1\otimes\ldots\otimes v_n=\frac 1 {n!}\sum_{\sigma\in S_n}(-1)^\sigma v_{\sigma(1)}\otimes\ldots\otimes v_{\sigma(n)}.\nn
\eea
The definition above is equivalent to the following 
\bea
det(B)=Tr_{1\ldots n}A_n B_1\ldots B_n,\nn
\eea
where  $B_k$ denotes an operator in $End(\C^n)^{\otimes n}\otimes A$
given by the fomrula 
\bea
B_k=\sum_{ij}1\otimes\ldots\otimes\underbrace{E_{ij}}_{k}\otimes\ldots\otimes 1\otimes B_{ij},\nn
\eea
the trace is taken on $End(\C^n)^{\otimes n}.$
\subsection{Quantum spectral curve}
We consider the same matrix $A$ as in the classical case but interpret its elements as generators of $U(\mathfrak{gl}_n)$
\bea
A=\sum_{ij}E_{ij}\otimes e_{ij}.\nn
\eea
Let us consider the generating series 
\bea
\label{gen_fun_sln}
P_Q(z,\p,\ep)&=&det\left(A z^{-1}+\xe
-\p Id\right)\nn\\
&=&\sum_k QI_k(z,\p,\ep)=\sum_k \ep^{k(k-1)/2}(QI_k^0(z,\p)+O(\ep)).
\eea
$QI_k(z,\p,\ep)$ are homogeneous in $(z^{-1},\p).$ \footnote{It is possible to speak about homogeneity due to the relation in Witt algebra $\p z- z\p=1$ which is homogeneous in $\p,z^{-1}.$}
The commutative algebra in $U(\mathfrak{gl}_n)$ is generated by the coefficients of 
\bea
QI_k^0(z,\p)=\sum_{l=0}^k QI_{k,i}z^{-i}\p^{k-i}.\nn
\eea
These operators are highest terms in $\ep$-expansion of the corresponding Gaudin Hamiltonians and hence commute. The general statement for the Gaudin model is proved in \cite{T04}, the case with the magnetic field is analyzed in \cite{CT}. 
{\Th The set 
$\{QI_{k,i}\}\subset U(\mathfrak{gl}_n)$ generates a commutative algebra $\mathcal{H}_q$ which quantizes the Poisson-commutative subalgebra in $S(\mathfrak{gl}_n)$ generated by $I_{k,i}.$
}
\subsection{Quantum invariance}
The subject of this part is the commutative subalgebra in $U(\mathfrak{b})$ quantizing the subalgebra of Hamiltonians for the generic Toda chain. We proceed by the same strategy as in the classical case - we will find invariants with respect to the action of the Borel subgroup $B$ on some localization of $U(\mathfrak{b})$.
\\
Let us define the decomposition problem in the quantum case. We always consider the decomposition of the Lie algebra
\bea\label{decomposition}
\mathfrak{sl}_n=b\oplus \mathfrak{so}_n
\eea
which transforms to the following one on the level of universal enveloping algebras:
\bea
U(\mathfrak{sl}_n)=\mathfrak{so}_n U(\mathfrak{sl}_n)\oplus U(\mathfrak{b})\nn
\eea
given by choosing the normal ordering. Let us denote by $a_+$ the projection of $a\in U(\mathfrak{sl}_n)$ to the second summand.

Let us demonstrate the specific invariance property of the quantum commutative family. The action of the group element $g\in B$ on $QI_{k,i}$ can be realized in terms of the action on the quantum Lax operator, which is the same as in the classical case:
\bea\label{quant_invar}
det\left(g^T A {g^T}^{-1} z^{-1}+\xe-\p\right)=det\left(A z^{-1}+{g^T}^{-1}\xe g^T-\p\right),
\eea
hence the question is reduced to the properties of the matrix $\xe.$
To prove (\ref{quant_invar}) we need the following lemma
{\Lem
Let $g\in GL_n$ and $L$ be an element of $Mat_n\otimes \mathfrak{R}$ with values in some associative algebra $\mathfrak{R}.$ Then $det(gLg^{-1})=det(L).$ 
}
~\\
{\bf Proof.~}
Let us use an alternative formula for the noncommutative determinant
\bea
det(L)=Tr_{1,\ldots,n}A_n L_1 L_2\ldots L_n.\nn
\eea
Then
\bea
det(gLg^{-1})&=&Tr_{1,\ldots,n}A_n g_1 L_1 g_1^{-1} g_2 L_2 g_2^{-1}\ldots g_n L_n g_n^{-1}\nn\\
&=&Tr_{1,\ldots,n}A_n g_1 g_2\ldots g_n L_1 L_2 \ldots L_n g_1^{-1} g_2^{-1}\ldots g_n^{-1}\nn\\
&=&Tr_{1,\ldots,n}g_1^{-1} g_2^{-1}\ldots g_n^{-1}A_n g_1 g_2\ldots g_n L_1 L_2 \ldots L_n \nn\\
&=&Tr_{1,\ldots,n}A_n L_1 L_2 \ldots L_n =det(L).
\eea
Here we have used the fact that $Tr(A B)=Tr( B A)$ for matrices with commuting entries
$[A_{ij},B_{kl}]=0;$ and the fact that the action of the symmetric group algebra commute with the diagonal linear group action on the tensor product of vector representation.
$\blacksquare$

\subsection{Quantum AKS lemma}
We have demonstrated that the group $B$ acts on $\Delta_k(z,\p)$ by a character $\chi_k(b)$ where $b\in B.$ Let us denote by the same letter the character on the Lie algebra $\chi_k(X)$ for $X\in \mathfrak{b}$ such that 
\bea
ad_X \Delta_{k,i}=\chi_k(X) \Delta_{k,i}.\nn
\eea
Let us also introduce a notation $\eta_a(m)$ for $a=\Delta_{k,i}$ and $m\in U(\mathfrak{b})$ such that
\bea
\label{char}
m a=a\eta_k(m).
\eea
For $m$ being a monomial $m=b_1 \ldots b_s$ 
\bea
\eta_k(m)=(b_1+\chi_k(b_1))\ldots (b_s+\chi_k(b_s)).\nn
\eea
{\Lem \label{lem_AKS}
Let us consider decompositions of the type (\ref{decomposition}) of two elements of our commutative algebra
\bea
QI_{k,i}&=&a=a_+ + a_-,\nn\\
QI_{l,j}&=&b=b_+ + b_-.\nn
\eea
Then 
\bea
a_+\eta_k(b_+)=b_+\eta_l(a_+).\nn
\eea
}
{\Rem This is an analog of the AKS lemma.}
~\\
{\bf Proof.~}
Let us use the commutativity condition
\bea
\label{eq1}
[a_-,b_-]&=&[a-a_+,b-b_+]=[a,b]+[a_+,b_+]-[a,b_+]-[a_+,b]\nn\\
&=&[a,b]+[a_+,b_+]-a b_+ + a\eta_k(b_+)- b \eta_l(a_+)+b a_+\nn\\
&=&[a_+,b_+]- a_- b_+ - a_+ b_+ + a_-\eta_k(b_+) + a_+ \eta_k(b_+) \nn\\
&-& b_- \eta_l(a_+)-b_+ \eta_l(a_+) + b_+ a_+ + b_- a_+.
\eea
The positive part of (\ref{eq1}) equals
\bea
0&=&a_+ \eta_k(b_+) - b_+ \eta_l(a_+).\nn
\eea
$\blacksquare$

We are interested in considering the localization of $U(\mathfrak{b})$ with the multiplicative set $S$ generated by $\{QI_{k,n-k}\}$ - the set of highest terms of all partial quantum characteristic polynomials. 
{\Lem $S$ is a right Ore set.}
~\\
Let us remind the Ore requirements. $S\subset A$ is a right Ore set if
\begin{enumerate}
 \item $\forall s\in S$ and $a\in A$  $\exists s'\in S$ and $a'\in A$ such that $ s a'= a s';$
 \item $\forall a_1,~ a_2\in A,~ \forall s\in S: (s a_1 =s a_2 ) \Rightarrow (\exists s'\in S:  a_1 s'=a_2 s').$
\end{enumerate}
The second condition is trivial because the algebra $U(\mathfrak{sl}_n)$ has no zero divisors. The first condition fulfills due to (\ref{char}). 
{\Th Let us consider the localization $loc_S U(\mathfrak{b}).$ Then the ratios $(QI_{k,i})_+/QI_{k,n-k}$ generate a commutative subalgebra:
\bea
[(QI_{k,i})_+/QI_{k,n-k},(QI_{m,j})_+/QI_{m,n-m}]=0\nn
\eea
which is a quantization of the classical generic Toda subalgebra.
}
~\\
{\bf Proof.~}
Let us use notations
\bea
a=QI_{k,i},~ b=QI_{m,j},~c=QI_{k,n-k},~d=QI_{m,n-m}.\nn
\eea
Then
\bea
a_+ c^{-1} b_+ d^{-1}-b_+ d^{-1} a_+ c^{-1}= a_+\eta_k(b_+)c^{-1} d^{-1}- b_+ \eta_m(a_+)d^{-1} c^{-1}.\nn
\eea
This is zero due to lemma \ref{lem_AKS} and the commutativity of $d$ and $c$ in both algebras.
$\blacksquare$

\section{Related topics}
\subsection{Drinfeld Zastava space}
The Zastava space was introduces as a completion of the space of maps from $\C P^1$ to the flag variety $\mathcal{F}_n$ of fixed degree 
\bea
\mathcal{M}_{\bar{d}}=\{f:\C P^1\rightarrow \mathcal{F}_n, c_1(f^* \mathcal{L}_i)=d_i\in H^1(\C P^1)\}\nn
\eea
where $\mathcal{L}_i$ is the ensemble of canonical line bundles on the flag variety $\mathcal{L}_i=V_i^{\wedge i}.$

We can consider a quite wider space of degree $d$ rational matrix-valued functions on $\C P^1$
\bea
L_d=\{F:\C P^1 \rightarrow Mat_{n}\}\nn
\eea
and realize the space $\mathcal{M}_{\bar{d}}$ factorizing by the Borel group action
\bea
f:\C P^1\stackrel{F}{\rightarrow}Mat_{n}\stackrel{/B}{\rightarrow} \mathcal{F}_n.\nn
\eea
The space $L_d$ has natural Poisson bracket given by the $R$-matrix structure
\bea
\{F_1(z),F_2(u)\}=[r(z-u),F_1(z)+F_2(u)] \nn
\eea
which allows to make the AKS reduction analogous to the generic Toda case. As a result one obtains an integrable system on the space $\mathcal{M}_{\bar{d}}$ due to the invariance of the constructed integrals with respect to the Borel group action. We suppose that the quantization technique of \cite{R} provides the same quantization for the generic Toda chain as those from our approach.

\end{document}